# Medical Information Embedding in Compressed Watermarked Intravascular Ultrasound Video

Nilanjan Dey[1] Suvojit Acharjee[1] Debalina Biswas[2] Achintya Das[3] Sheli Sinha Chaudhuri[1]

**Abstract** – In medical field, intravascular ultrasound (IVUS) is a tomographic imaging modality, which can identify the boundaries of different layers of blood vessels. IVUS can detect myocardial infarction (heart attack) that remains ignored and unattended when only angioplasty is done. During the past decade, it became easier for some individuals or groups to copy and transmits digital information without the permission of the owner. For increasing authentication and security of copyrights, digital watermarking, an information hiding technique, was introduced. Achieving watermarking technique with lesser amount of distortion in bio-medical data is a challenging task. Watermark can be embedded into an image or in a video. As video data is a huge amount of information, therefore a large storage area is needed which is not feasible. In this case motion vector based video compression is done to reduce size. In this present paper, an Electronic Patient Record (EPR) is embedded as watermark within an IVUS video and then motion vector is calculated. This proposed method proves robustness as the extracted watermark has good PSNR value and less MSE.
Keywords: Watermark, IVUS, EPR, Motion vector

## I. INTRODUCTION

Intravascular ultrasound is a medical imaging technique which uses an ultrasound probe positioned at the tip of a coronary catheter. This methodology uses the ultrasound technology to visualise inner walls of the blood vessels (endothelium), inside out. Intravascular ultrasound is basically a tomographic imaging modality performed during coronary angiography. The most frequently imaged part of the body, by the IVUS [1, 2, 3, 4], happens to be the arteries of the heart. Accumulation of calcium in the epicardial coronary artery is one of the many factors that lead to heart attack. Atheromatous plaque results from deposition of macrophage cells or debris and lipid (cholesterol) inside the artery walls. This deposition over the time subsequently leads to narrowing of the arteries and heart attack. Angiography fails to detect plaque and herein lies the advantage of IVUS. IVUS helps to get a clear view of not only the lumen (opening), but also the atheroma (membrane/cholesterol loaded white blood cells) concealed within the wall. Thus IVUS aids in detecting a vital reason for myocardial infarction (heart attack) that remains ignored and unattended when only angioplasty is in use.

Intravascular ultrasound is capable of determining the relationship between plaque and vessel wall in real time throughout the coronary arterial tree. This exactly defines the quantity as well as the distribution of calcium within the vessel wall. Intravascular ultrasound is able to classify different plaque substructures and also can help to clarify the ambiguous angiogram as well as delineate the exact nature of luminal encroachment. From treatment point of view, the identification of calcification patterns, particularly those on the superficial intimal surface, can alert the operator to change the compliance prior to definitive therapy.

IVUS can measure the diameter of stent, stent symmetry and stent expansion. Stent being made by metals reflects a strong layer of ultrasound. Segments of blood vessels, positive and negative atheroma development can also be measured by IVUS. Thus, IVUS is capable of imaging the inside of vessels, measuring the plaque, detecting exact vessel composition.

During the past decade, with the expansion of information digitization and internet, digital media progressively dominated the conventional analog media. As the flip side of the coin, there is an associated disadvantage of digitization and internet. It is also becoming easier for some individuals or groups to copy and transmit digital information without the permission of the owner. The digital watermark is then introduced to solve this problem. Watermarking is a branch of information hiding which is used to hide proprietary information in digital media like digital images and signals. The easiness with which digital content can be exchanged over the Internet has created copyright violation issues. Copyrighted material can be easily exchanged over peer-to-peer networks, and this has caused major concerns to those content providers who produce these digital contents.

[1]Electronics and Telecommunication Engineering Dept., Jadavpur University, Kolkata, West Bengal, India.
[2]Computer Science & Engineering Dept., JIS College of Engineering, Kalyani, West Bengal, India.
[3]Electronics & Communication Engineering Dept., Kalyani Govt. Engineering College, Kalyani, India.

The use of internet is the most important media of globalization. In the present mechanized age, globalization has influenced the medical field also. Intelligent exchange of bio-medical images and bio-signals amongst hospitals and diagnostic centres for mutual availability of therapeutic case studies needs highly efficient and reliable transmission. Watermarking technique is also being used in medical field to protect bio-medical information while transmitting through the wireless media. Watermark [5,6,7] is added ownership to increase the level of security and to verify authenticity. Patients' information (Electronic Patient Record), logo of hospitals or diagnostic centres can be added in the bio-medical data as watermark to prove the intellectual property rights. Addition of watermark in a medical signal or image can cause distortion. As all the bio-medical images and signals convey information required in diagnosis of diseases, any kind of distortion is not acceptable. But for authenticity and security of the information a little amount of distortion can be overlooked. So achieving watermarking technique with lesser amount of distortion in bio-medical data is a challenging task.

Watermark can be embedded into an image or in a video. Video generally refers to sequence of images. Video files contain a huge amount of data. Hence huge amount of storage is required to store a video sequence and a wide bandwidth required to transmit a video sequence. But it is not feasible to provide a huge storage or wide bandwidth for every video sequence. Video compression is the most useful way for reducing the video size thus reducing the required amount of bandwidth to transmit the video. Video compression is actually a process in which the amount of data used to represent the video is reduced to meet the bandwidth requirement of the network or space requirement.

The major concern is that the watermark embedded with the video must retain its quality. In case of video compression the quality of the reconstructed video is the main matter of concern. In most of the application fields only lossless video compressions are allowed. As the bio-medical videos contain necessary information to diagnose several diseases, lossy video compression is not acceptable in the medical field. But in few applications such as TV transmission a certain amount of information loss is allowed. This type of compression is referred to as lossy compression. By eliminating the redundancy present in an image sequence video compression can be achieved.

To remove the temporal redundancy present in a video sequence the motion present in the video sequence need to be found out. Motion vector estimation is the most important part of any video compression scheme. In this present age the use of multimedia and internet has become very common. Hence the issues regarding downloading video streams and storing them in various media have become very important, as they need a lot of memory. The ISO Moving Picture Experts Group (MPEG) video coding standards is for compressed video storage in physical media like hard disk, and International Tele-communications Union (ITU) [8] addresses the real time point to point or multipoint communication over a network. In both the cases the entire compression and decompression process is largely the same. Fig. 1 shows the block diagram for video compression process.

The most computationally expensive part in the compression process is the Motion Estimation. In motion compensated coding, a uniform motion model is assumed. It is assumed that change in successive frames is due to translational movement of the objects. Motion Estimation examines the movement of objects in sequence to try to obtain the vectors representing the estimated motion. Encoder side estimates the motion of the current frame with respect to previous frame. A motion compensated image of the current frame is then created. Motion vector is then transmitted to decoder. Decoder reverses the whole process and creates a full frame. This way motion compensation uses the knowledge of object motion to achieve data compression. Motion Estimation algorithms have few assumptions as follows:

1) Objects move in translation in a plane that is parallel to the camera plane, i.e., the effects of camera zoom, and object rotations are not considered.

2) Illumination is spatially and temporally uniform.

3) Occlusion and disocclusion effects are neglected.

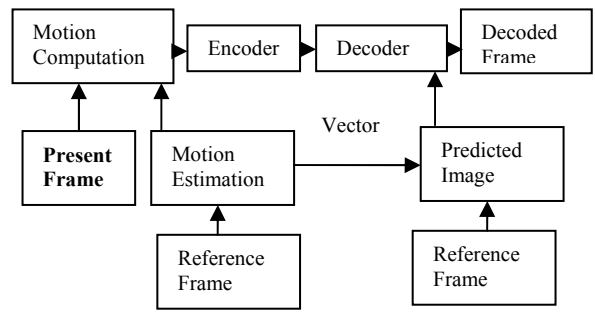

Fig.1. Block Diagram for Video Compression

The motion compensated coding can be split into three stages.

At first, calculation for motion vector estimation, either for a pixel or for a group of pixels, is done. This step is referred to as motion analysis stage.

Then the present frame is predicted using the previous frame and the predicted motion vector. The difference between actual present frame and predicted present frame is calculated. This is the prediction error.

The last part is encoding. In this stage the prediction error and the predicted vector are encoded.

There are two main techniques for motion vector estimation.

Pel-Recursive Algorithm.
Block Matching Algorithm

Pel-Recursive Algorithm (PRA) introduced by Netravali and Robbins, defines motion vector for every pixel which comprises the location of the pixel of the current frame in the previous frame. In Block Matching Algorithm (BMA), it is assumed that every pixel within a macro block has the same motion activity and produces one motion vector for each macro block. The main idea behind block matching is to divide the current frame into a number of macro blocks of fixed size and create a motion vector which comprises the location of the macro block of the current frame in the previous frame. Usually the macro block is taken as a sequence of 16 pixels and search area is up to 7 pixels on all fours sides of the corresponding macro block in previous frame. The matching of one macro block with another is based on the output of a cost function. The macro block that results in the least cost is the one that matches the closest to current block. There are various cost functions, of which the most popular and less computationally expensive is Mean Absolute Difference (MAD) given by equation (1). Another cost function is Mean Squared Error (MSE) given by equation (2).

$$MAD = 1/N^2 \sum_{i=0}^{N-1} \sum_{j=0}^{N-1} | cufra(i,j) - refra(i,j) | \quad (1)$$

$$MSE = 1/N^2 \sum_{i=0}^{N-1} \sum_{j=0}^{N-1} (cufra(i,j) - refra(i,j))^2 \quad (2)$$

N= Size of the macro block.

*Cufra (i, j)* = Intensity value for the current frame.

*refra (i, j)* =Intensity value for the reference frame

Peak Signal to Noise Ratio (PSNR) characterizes the motion compensated image that is created by using motion vectors and macro blocks from the reference image.

In this paper after embedding the watermark within an IVUS video, the motion vector for that video signal has been found out. In the next step the video is compressed using motion vector estimation technique. Thus a good quality video of smaller size comparative to the original video is generated.

## II. METHODOLOGY

### A. Motion Vector Estimation

Motion vector estimation is the most computationally expensive process. Many algorithms have been developed to reduce the computational complexity of motion vector estimation. In the initial days Netravali and Robbins [9] in the year 1979 first developed a pel-recursive algorithm which defines motion vector for every pixel. Later Bargman [10] modified the algorithm. Walker and Rao[11] later defined a new variable step size for pel-recursive algorithm. In new fast motion vector estimation algorithm the difference in intensity of different pixels is used as search criteria.

Block matching algorithm for motion estimation is accepted in all the video coding standards proposed till date. It is easy to implement in hardware and real time motion estimation and prediction is possible. Exhaustive search (Full Search) is the most time and computationally expensive algorithm. Later the other algorithms implemented to reduce the computation cost are Three Step Search (TSS) [12], Four Step Search (FSS) [13], New Three Step Search (NTSS) [14], Diamond Search (DS) [15], and Adaptive Rood Pattern Search (ARPS) [16] etc.

In block matching algorithms square search pattern provides the best result. In this algorithm one motion vector is estimated for every block. Generally, the standard size of the blocks for most of the block matching algorithms is 16x16 or 8x8 non-overlapping blocks (macro blocks). As more vectors are needed to compute, selection of smaller block size leads to poor time complexity. If the blocks are too small, probably our proposed algorithm will consider the blocks as noise. But selection of too large blocks affects the quality, as the motion matching is most likely less correlated. In our proposed method, the block size has been taken to be 4x4 which is optimum for trade-off between compression efficiency and computational complexity.

The image is divided into blocks and the mean of the luminance intensity of each block is calculated. For every block one point is defined and is stored into a different matrix.

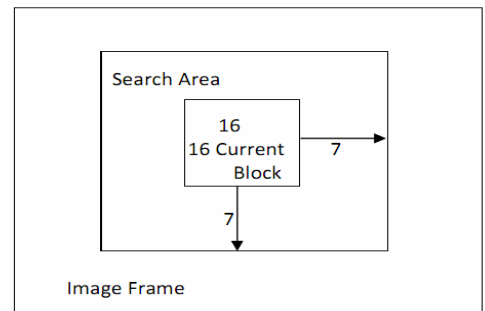

Fig. 2. Block Matching of a 16x16 macro Block within a search area of 7 pixels

## III. PROPOSED METHOD

After dividing the original IVUS video into frames, watermark is added.

### A. Watermark Embedding

Step 1. The whole image is divided into 8x8 blocks.

- Step 2. The watermark image contains only gray level 255 and 0. So the pixel values are changed. Pixels of values 255 are changed to 20 and pixels of values 0 are changed to 10.
- Step 3. Watermark image is divided into 2X1 block.
- Step 4. The blocks of watermark image are then put into the 8x8 blocks of the video frame. In each block of the frame the block of watermark image is inserted in the first two positions. This process will continue until all the blocks of the watermark image are placed into the video frame.

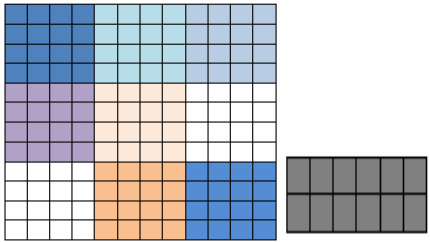

Fig. 3. Original Frame of size 12x12 (Original Frame is divided into 4x4 blocks and the watermark image of size 2x6 is divided into 1x2 blocks.)

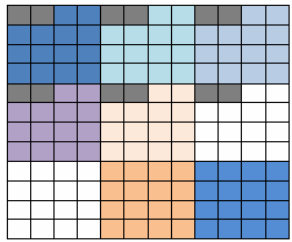

Fig. 4. Watermarked frame after embedding

B. *Motion vector Calculation*

- Step 1. The current and reference frame is divided into 4x4 blocks.
- Step 2. Means for the image blocks in the current and reference frames are calculated and stored in different matrices.
- Step 3. The block position from the current frame whose best match needs to be found is picked up. The equivalent point of the block in the current mean matrix is found out. The block in the same position as the center block in the reference frame and the same point as center search point in the reference mean matrix is set.
- Step 4. A search window from the reference mean matrix of a length 9x9 is created on all the four sides of the center search point. The center search point will remain in the center of the search window. So it will cover four blocks on all the sides of the center block.
- Step 5. The difference between the center search point in the current mean matrix and every point in the search window is found out.
- Step 6. The point from the search window with minimum difference is found out. The distance of the point from the center point is also determined.
- Step 7. Multiplication between the difference and the block size is carried out to get the actual motion vector. So here the differences need to get multiplied by 4.
- Step 8. Calculated motion vector is saved for transmission.

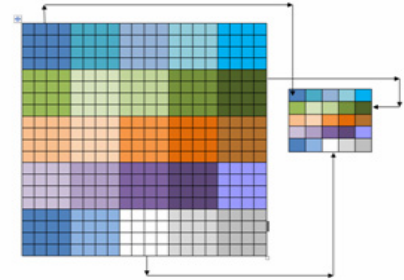

Fig. 5. The image is divided into blocks and calculated mean for every block is stored in a different mean matrix.

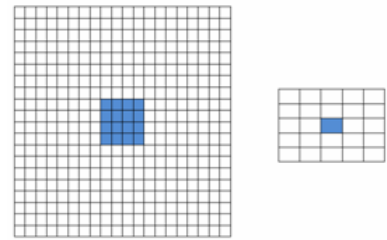

Fig. 6. The Block whose best match needs to be found and point representing the block in the mean matrix

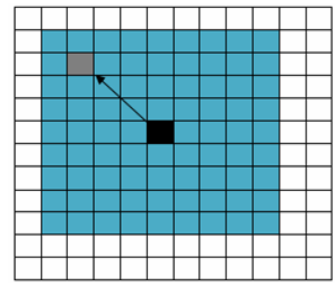

Fig. 7. Search window, center point and the point with minimum difference

V. RESULT & DISCUSSIONS

The proposed method is simulated using a medical video [17, 18]. Length of the video is 4 seconds and number of frames present in the video is 20. MATLAB 7.0.1 software is extensively used for the study of video. Concerned images obtained in the result are shown in Figure 8 through 15.

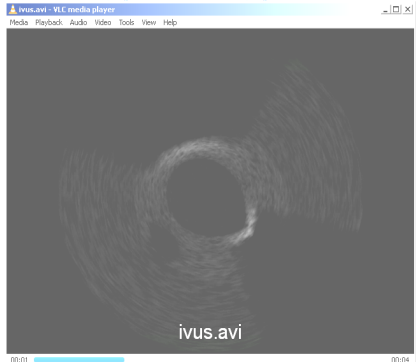

Fig. 8. Screenshot of the original video

The watermark image which has to be embedded within the video is of length 31X64. It consists of two gray level 0 and 255.

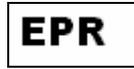

Fig. 9. The watermark image

The watermark is embedded within every frame of the video as described in the proposed method. After embedding the watermark, the motion vector estimation and video compression for the video process is done. After compression, a new file with *.s2f extension is generated which consists the video in a compressed format. To view the video as an uncompressed one, the video has to be recovered from the file. The recovered video has a good PSNR response and MSE response which reflects that the recovered video retains the original quality of video.

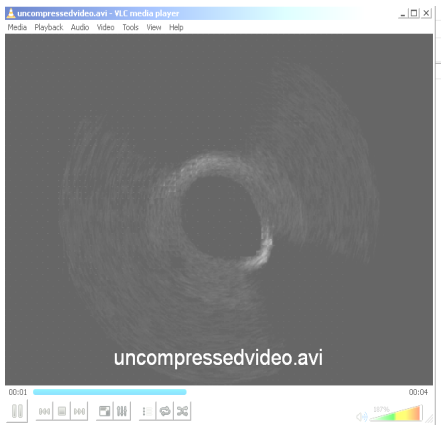

Fig. 10. Screenshot of the recovered video from the compressed file

### A. Peak Signal to Noise Ratio (PSNR)

It measures the quality of a watermarked signal. This performance metric is used to determine the perceptual transparency of the watermarked signal with respect to original signal:

$$PSNR = \frac{XY \max_{x,y} P_{x,y}^2}{\sum_{x,y}(P_{x,y} - \overline{P}_{x,y})^2} \quad (3)$$

Where, M and N are number of rows and columns in the input signal, $P_{x,y}$ is the original Signal and $\overline{P}_{x,y}$ is the watermarked Signal. The PSNR of every frame in the original video and recovered video is given in the following graph. Here after every 6th frame one original frame is sent. So after every 6th frame PSNR becomes zero.

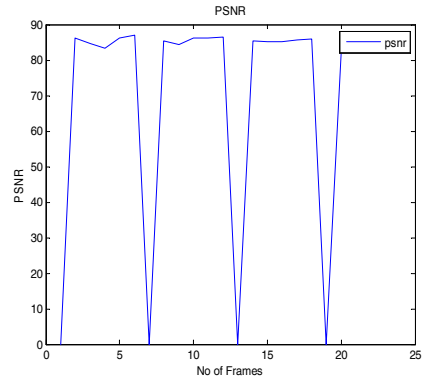

Fig. 11. PSNR of the original and recovered video

MSE of the recovered video frames compared with original video frames is minimum as shown in the below graph.

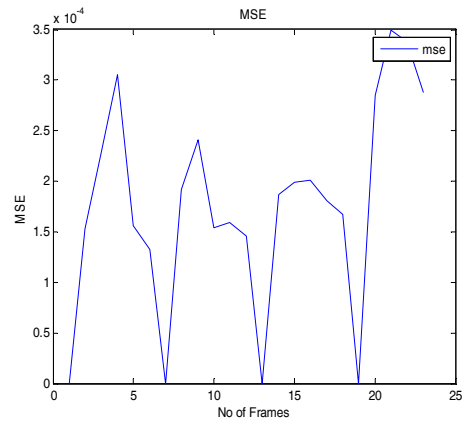

Fig. 12. MSE of the original and recovered video.

In the figure given below the 3rd frame of the original video and recovered video has been shown.

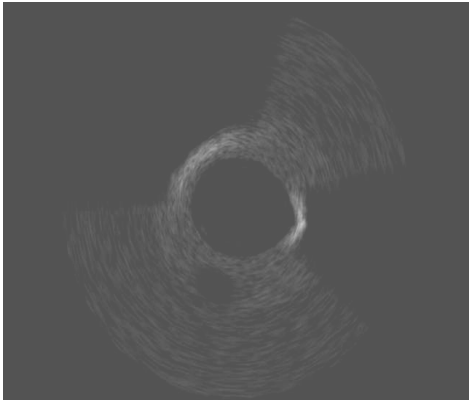

Fig. 13. Third frame of the original video sequence

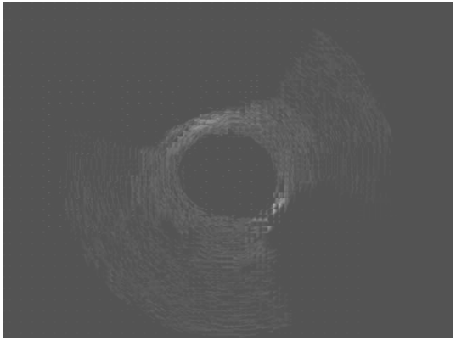

Fig. 14. Third frame of the Recovered Video sequence

Fig 15: Recovered watermark from the video

### B. Correlation Coefficient

After secret image embedding process, the similarity between the original signal x and watermarked signal x' is measured by the standard correlation coefficient as follows:

$$C = \frac{\sum_m \sum_n (x_{mn} - x')(y_{mn} - y')}{\sqrt{\left(\sum_m \sum_n (x_{mn} - x')^2\right)\left(\sum_m \sum_n (y_{mn} - y')^2\right)}} \quad (4)$$

where y and y' are the discrete wavelet transforms of x and x'.

Table 1: Correlation of the first 10 frames of the original Watermark image and recovered Watermark image

| Correlation Table | |
|---|---|
| Correlation between original Watermark and Recovered Watermark | Value |
| Frame 1 | 0.91 |
| Frame 2 | 0.89 |
| Frame 3 | 0.889 |
| Frame 4 | 0.88 |
| Frame 5 | 0.88 |
| Frame 6 | 0.89 |
| Frame 7 | 0.91 |
| Frame 8 | 0.89 |
| Frame 9 | 0.88 |
| Frame 10 | 0.88 |

### C. *Structural similarity index metric (SSIM):*

Structural similarity index metric (SSIM) is based on the structural information of the image. It provides a good measure for different kinds of images, from natural scenes to medical images. SSIM is a recent proposed approach for image quality assessment. SSIM index is capable of measuring similarities between two images. The SSIM is designed to improve on traditional metrics like PSNR and MSE, which have proved to be inconsistent with human eye perception. If two non negative images are placed together,

Mean intensity can be represented by the following equation:

$$\mu_x = 1/N \sum_{i=1}^{N} x_i \quad (5)$$

Standard deviation can be represented by the following equation:

$$\sigma_x = \left(\frac{1}{N-1} \sum_{i=1}^{N} (x_i - \mu_x)^2\right)^{1/2} \quad (6)$$

Contrast comparison c(x,y) - difference of $\sigma_x$ and $\sigma_y$

$$c(x,y) = \frac{2\sigma_x \sigma_y + C_2}{\sigma_x^2 + \sigma_y^2 + C_2} \quad (7)$$

Luminance comparison can be defined by the following equation:

$$l(x,y) = (2\mu_x \mu_y + C_1)/\mu_x^2 + \mu_y^2 + C_1 \quad (8)$$

where C1, C2 are constant.

Structure comparison is conducted s(x,y) on these normalized signals $(x-\mu_x)/\sigma_x$ and $(y-\mu_y)/\sigma_y$,

$$S(x,y) = f(l(x,y), c(x,y), s(x,y)) \quad (9)$$

$$SSIM(x,y) = [l(x,y)]^\alpha \cdot [C(x,y)]^\beta \cdot [S(x,y)]^\gamma \quad (10)$$

$$SSIM(x,y) = \lfloor(2\mu_x\mu_y + C_1)(2\sigma_{xy} + C_2)\rfloor / (\mu_x^2 + \mu_y^2 + C_1)(\mu_x^2 + \mu_y^2 + C_2) \quad (11)$$

α, β and γ are parameters used to adjust the relative importance of the three components.

Now we can find the SSIM index table below:

Table 2: SSIM Index of the first 10 frames of the original Watermark image and recovered Watermark image

| SSIM Index Value Table | |
|---|---|
| SSIM Index between original Watermark image and recovered Watermark image obtain from Watermarked Video | Value |
| Frame 1 | 0.8387 |
| Frame 2 | 0.6944 |
| Frame 3 | 0.6944 |
| Frame 4 | 0.6944 |
| Frame 5 | 0.6944 |
| Frame 6 | 0.6944 |
| Frame 7 | 0.8387 |
| Frame 8 | 0.6944 |
| Frame 9 | 0.6944 |
| Frame 10 | 0.6944 |

VI. CONCLUSION

In medical field, exchange of biomedical data through the internet has increased a lot, whereas the level of security and authentication of data should also be kept in mind. In this paper, an EPR is watermarked in an IVUS video. As video data consists of a series of images and needs huge storage and wide bandwidth, therefore video data needs to be compressed. Here motion vector is calculated to reduce storage. For finer results the image is divided into 4x4 blocks. Selection of 8x8 blocks for watermarking 2 watermark bits, affects only 3.125 % of the overall block which is acceptable. The efficacy of the method is proved by the PSNR and MSE performance of the recovered video. Correlation value and SSIM are calculated to show the similarity between original and recovered watermark image.